\documentclass[twocolumn,showpacs,preprintnumbers,amsmath,amssymb]{revtex4}

\input epsf

\newcommand{\beq}{\begin{eqnarray}}% can be used as {equation} or {eqnarray}
\newcommand{\eeq}{\end{eqnarray}}

\def\be{\begin{equation}}
\newcommand{\bel}[1]{\be\label{#1}}
\def\ee{\end{equation}}
\newcommand{\eref}[1]{(\ref{#1})}
\newcommand{\Eref}[1]{Eq.~(\ref{#1})}
\newcommand{\rem}[1]{}

\def\bit{\begin{itemize}}
\def\eit{\end{itemize}}

\def \be {\begin{equation}}
\def \ee {\end{equation}}
\def \bea {\begin{eqnarray}}
\def \eea {\end{eqnarray}}

\def\rd{$\rho$--dominance}
\def\rcu{$\rho$--coupling universality}

\newcommand{\matt}[1]{{}}%{\small\it{#1}}}
\newcommand{\cut}[1]{{}}%\small\bf{#1}}}
\newcommand{\move}[1]{{}}%\small\underline{#1}}

\begin{document}
\preprint{UW/PT 05-03,
hep-th/0501197}

\date{January 20, 2005}
\title{On the Couplings
of the Rho Meson in AdS/QCD}

\author{Sungho Hong, Sukjin Yoon,
and Matthew J. Strassler\\
Department of Physics,
P.O Box 351560, University of Washington, 
Seattle, WA 98195\\
}

\begin{abstract}We argue that in generic AdS/QCD models (confining gauge
theories dual to string theory on a 
weakly-curved background), the couplings $g_{\rho HH}$ of
any $\rho$ meson to any hadron $H$ are quasi-universal, lying within a
narrow band near $m_\rho^2/f_\rho$.  The argument relies
upon the fact that the $\rho$ is the lowest-lying state created by a
conserved current, and the detailed form of the integrals
which determine the couplings in AdS/QCD.  Quasi-universality holds even when rho-dominance is violated. The argument
fails for all other hadrons, except for the lowest-lying spin-two
hadron created by the energy-momentum tensor.  Explicit examples are
discussed.
\end{abstract}

\maketitle

%  READ SAKURAI'S WORK

%  CHECK THE G-HATS

%  CHECK THE ARGUMENT ABOUT THE DERIVATIVE OF THE RHO WAVE FUNCTION

%  DEAL WITH THE MUS

%  FIGURES SHOWING FIT TO RCU

The couplings of the $\rho$ meson to pions and to nucleons are
remarkably similar, $g_{\rho\pi\pi}\sim g_{\rho NN}$ \cite{oldwork}.  Is
this accidental or profound?  Most other couplings of the $\rho$ 
cannot be easily
measured; even the coupling $g_{\rho\rho\rho}$ is unknown. 
Lattice gauge theory at large number of colors $N$, where
hadrons are more stable and the quenched approximation is valid, could
potentially be used to obtain additional information, but there
have been few if any efforts in this direction.  In the absence of
constraints from data or from numerical simulation, the issue of
whether the $\rho$ has universal couplings to all hadrons has been
left to theoretical speculation.

In this letter we will reexamine the long-standing ``\rcu''
conjecture \cite{oldwork}.  We view the conjecture as having two parts: 
(a) the $\rho$ has
universal couplings, and (b) the universal coupling is equal to
$m_\rho^2/f_\rho$.  There is very little reason to expect this
conjecture to be exact in QCD, but even attempts to explain its
approximate validity have relied upon particular arguments which
themselves are open to question.  We will reexmaine this
web of arguments in AdS/QCD.  AdS/QCD offers us the
opportunity to compute all the couplings of an infinite
number of hadrons in a four-dimensional confining gauge theory.  This
makes it an ideal setting for testing theoretical
arguments concerning the properties of hadrons.  We will 
examine the AdS/QCD calculation of the $\rho$'s couplings, 
and make estimates that show they lie in a narrow band near
$m_\rho^2/f_\rho$.  We will then check that this is actually true in
explicit models.

We will
consider the form factor $F_H(q^2)$ for a hadron $|H\rangle$ with
respect to a conserved spin-one current. 
The same current, applied to the vacuum, creates a set of spin-one
mesons $|n\rangle$, where $n=0,1,\dots$; we will refer to the $n=0$
state as the ``$\rho$''.  At large $N$, a form factor for a hadron $H$
can be written as a sum over vector meson poles,
\begin{equation}\label{Ffg}
\quad F_{H} ( q^2 ) = \sum_n \frac{f_n g_{n HH}}{q^2 + m_n^2},
\end{equation}
where $m_n$ and $f_n$ are the
mass and decay constants of the vector meson $|n\rangle$, 
and $g_{nHH}$ is its coupling to the hadron
$H$.  (Henceforth we will use both subscript-$\rho$ and subscript-$0$
to denote quantities involving the $\rho$.)  
 Charge conservation
normalizes the form factor exactly: we factor out the total charge of
the hadron $H$ in our definition of $F_H(q^2)$, so that 
\bel{Fnorm}
F_H(0)= 1 =  \sum_n \frac{f_n g_{n HH}}{m_n^2} \ .
\ee

A classic argument in favor of \rcu\ rests
upon an assumption, \rd, which is supported to some degree
by QCD data.  
There is some ambiguity in the terminology,
but by our definition, ``\rd'' means
that the $\rho$ gives by far the largest
contribution to the form factor at small $q^2$:
\bel{rddefn}
{f_0g_{0HH}\over q^2 + m_0^2} \gg {f_n g_{nHH}\over q^2 + m_n^2}
\ \ \ (|q^2| \alt  m_0^2\ , \ n>0)
\ee
Combined with \eref{Fnorm}, this condition implies, subject to certain
convergence criteria, that
\bel{Fnormrd}
1 = {f_0g_{0HH}\over  m_0^2} +\sum_{n=1}^\infty {f_n g_{nHH}\over m_n^2}
\approx {f_0g_{0HH}\over m_0^2}
\ee
which in turn 
proves $g_{0HH}\approx m_0^2/f_0$ for all $H$ --- in short, \rcu.  

However, the convergence conditions for the sums over $n$ are not
necessarily met.  Convergence cannot, of course, be checked using
data.  Meanwhile, \rd, though a sufficient condition for \rcu, is
clearly not necessary.  The individual terms in the sum in \eref{Fnormrd}
could be large and of alternating sign, and still permit \rcu.  We
will see later that this does happen in explicit AdS/QCD
examples.

A second and qualitatively different argument 
treats the $\rho$ meson as a gauge boson \cite{oldwork}, with the
hope that the broken four-dimensional gauge symmetry might assure
\rcu.  Hidden local symmetry \cite{hls} is a consistent
formulation of this idea,
but does not have exact \rcu\ as a consequence
(except, possibly, in a limit when the $\rho$
becomes massless relative to all other vector mesons,
as proposed in \cite{vm}.)
Instead, it is related  \cite{SonSteph} to a deconstructed
version of AdS/QCD, and is subject to the arguments
given below.

In AdS/QCD, where gauge theories with 't Hooft coupling $\lambda=g^2N$
($g$ the Yang-Mills coupling, $N$ the number of colors) are related
\cite{AdS/CFT} to string theories on spaces with curvature $\sim
1/\sqrt\lambda$, these issues take on a new light.  Although the
$\rho$ cannot be treated as a four-dimensional gauge boson, a $\rho$
meson in AdS/QCD is the {\it lowest cavity mode} of a {\it
five}-dimensional gauge boson. In the limit $\lambda\to\infty$, the
$\rho$ meson, along with the entire {\it tower} of vector mesons ---
the remaining cavity modes of the five-dimensional gauge boson ---
becomes massless.  More precisely, the vector mesons become
parametrically light compared to the inverse Regge slope of the
theory, and thus to all higher-spin mesons.  However, the universal
properties of the {\it five}-dimensional gauge boson do not imply
\rcu; they simply ensure that the corresponding global symmetry charge
is conserved and that $F(q^2\to 0) = 1$.

Nevertheless, the AdS/QCD context offers a new and logically distinct
argument for {\it generic and approximate} \rcu, as suggested by
\cite{haduniv}.  This argument does not rely upon the
existence of a limit in which \rcu\ is exact, and indeed there is no
such limit.  Even at infinite $N$ and/or infinite $\lambda$, the
couplings of the $\rho$ always remain quasi-universal.

We will consider the
string-theoretic dual descriptions of four-dimensional confining gauge
theories which are asymptotically scale-invariant in the ultraviolet.
For large $\lambda$ and $N$ the dual description reduces to ten-dimensional
supergravity, on a space with four-dimensional Minkowski coordinates
$x^\mu$, a ``radial'' coordinate $z$, and five compact coordinates
$\Omega$.  The coordinates can be chosen to put the metric in the form
$ds^2 = e^{2A(z)} \eta_{\mu\nu}dx^\mu dx^\nu + R^2 dz^2/z^2 + R^2
d\hat s_\perp^2$; here $R^2 d\hat s_\perp^2$ is the metric on the five compact
directions, and $R\sim \lambda^{1/4}$ is the typical curvature radius
of the space.  The coordinate $z$ corresponds to $1/{\rm Energy}$ in
the gauge theory.  The ultraviolet of the gauge theory corresponds to
$z\to 0$; the gauge theory is nearly scale-invariant in this region,
and the metric correspondingly is that of $AdS_5$ (with
$e^{2A(z)}=R^2/z^2$) times a five-dimensional compact space $W$.
% with
%metric $ds^2_W=R^2 d\hat s^2_W$.  
The infrared, where confinement
occurs, is more model-dependent, but generally the radial coordinate
terminates, at $z= z_{max}\sim 1/\Lambda$ 
\cite{witten,ikms}.  Importantly, the metric
deviates strongly from $AdS_5$ --- scale invariance is strongly broken
in the gauge theory --- only for $z$ on the order of $z_{max}$.

The form factor of a hadron $H$ associated to a conserved current
$J^\mu$ can be computed easily in AdS/CFT.  
A conserved current in the gauge theory corresponds to a
five-dimensional gauge boson, which may arise as a gauge boson in ten
dimensions (or a subspace thereof) or as the dimensional reduction of
a mode of a ten-dimensional graviton.  This gauge field satisfies
Maxwell's equation (or the linearized Yang-Mills equation) in the
five-dimensional space, subject to Neumann boundary conditions
at $z\to z_{max}$.  For each four-momentum $q^\mu$ there is a
corresponding five-dimensional non-normalizable mode
$\Psi(q^2,z)e^{iq\cdot x}$ of the gauge boson.  (The mode will also
have some model-dependent structure in the remaining five dimensions,
but this structure always factors out because the
gauge boson arises from a symmetry --- for examples see
\cite{DIS,haduniv}.)  
The form factor is obtained by integrating this
mode against the current built from the incoming and outgoing hadron
$H$.  We take $H$ to be spin-zero; the generalization to higher spin
is straightforward.  From the ten-dimensional wave function of the
incoming hadron of momentum $p$, $\Phi_H(x,z,\Omega)= e^{ip\cdot
x}\phi(z,\Omega)$, and the corresponding mode for the outgoing hadron
with four-momentum $p'$, we construct the current $J_H(x,z)$: 
\be
J^\mu_H = -i\int\ R^5 d^5\Omega \ \sqrt{\hat{g}_\bot}\ \left\{
\Phi_H^* \overleftrightarrow\partial^\mu
\Phi_H' \right\} \ .  
\ee
Here $\hat g$ is the metric of $W$.
Equivalently, $J^\mu_H(z) = (p+p')^\mu e^{i(p-p')\cdot x}\sigma_H(z)$
where
$$
\sigma_H(z) = \int \ 
R^5 d^5\Omega \ \sqrt{\hat{g}_\bot} \
 \phi_H^*(z,\Omega) \phi_H(z,\Omega) \geq 0  \ .
$$
The hadron wave functions are normalized to unity:
\begin{equation}\label{Hnorm}
\int \ dz\ \mu_H(z) \ \sigma_H(z) = 1 \ .
\end{equation}
where $\mu_H = e^{2A(z)}R/z$ (for a spin-zero hadron.)
In terms of $\sigma_H$, the form factor then reduces to
\begin{equation}
  \label{eq:formfactor}
F_{H} ( q^2 ) = g_5 \int \ dz\ \mu_H(z) \ 
\Psi ( q^2,z ) \sigma_H(z)
 \ .
\end{equation}
As $q^2\to 0$, the non-normalizable mode $\Psi(q^2,z)$ goes to
a constant, namely $1/g_5$; then \Eref{eq:formfactor} becomes
equal to \Eref{Hnorm}, enforcing the condition $F_H(0)= 1$.

Meanwhile, the $\rho$, as the lowest-mass state created by the
conserved current, appears in AdS/QCD as the lowest normalizable
four-dimensional cavity mode of the same five-dimensional gauge boson.
The coupling $g_{0HH}$ is computed in almost the same way, by
integrating the $\rho$'s ten-dimensional wave function $\varphi_0$
(which is trivial in the five compact dimensions) against the current
of the ten-dimensional wave function $\Phi_H$:
\begin{equation}
  g_{0HH} = g_5 \int dz\ \mu_H(z)
\varphi_0(z) \sigma_H(z) \ .
\label{eq:THO} 
\end{equation}
Couplings $g_{nHH}$ for the $n^{th}$ vector meson are computed
by replacing $\varphi_0(z)$ with the $n^{th}$ mode function
$\varphi_n(z)$.

The function $\varphi_0(z)$, as the lowest normalizable solution of
a second-order differential equation, is typically structureless and
has no nodes. It can be chosen to be positive definite.  On general
grounds it grows as $z^2$ at small $z$ (near the boundary) until
scale-invariance is badly broken, in the region $z\sim z_{max}$.  Also,
because the $\rho$ is a mode of a conserved current, it must satisfy
Neumann boundary conditions (so that $\Psi(0,z) = 1/g_5$ is an
allowed 
solution.)  Generically it will not vanish
at $z_{max}$.   We therefore expect that in the region $z\sim z_{max}$
the wave function $\varphi_0(z)$ has a finite typical
size
$\hat\varphi_0$.

We will now use these properties to make some estimates of the above
integrals.   The 
normalizable and non-normalizable modes are related by
\bel{relate}
\Psi(q^2,z) = \sum_n {f_n\varphi_n(z)\over {q^2+m_n^2}} \ .
\ee
Meanwhile,
the $\rho$ meson is normalized:
\bel{rhonorm}
1 = \int \ dz \ \mu_0(z) \varphi_0(z)^2  \sim \hat\mu_0 \hat\varphi_0^2 
\delta z \ ,
\ee
where $\mu_0(z) = R/z$ is a slowly-varying measure factor, 
$\hat\mu_0$ is the typical size of $\mu_0$ in the region
$z\sim z_{max}$,
and $\delta z\sim z_{max}/2$ is
the region over which $\varphi_0(z)\sim \hat\varphi_0$.
Since
$\Psi(0,z)=1/g_5$, applying \Eref{rhonorm}
to \Eref{relate} and using orthogonality of the modes $\varphi_n$ implies
\bel{fm2}
f_0/m_0^2 = {1\over g_5}
\int \  dz \ \mu_0(z) \varphi_0(z) \sim {1\over g_5}\hat\mu_0 \hat \varphi_0
\delta z \ .
\ee
%The ratio of  Eqs.~\eref{rhonorm} and \eref{fm2} gives
%$
%\hat \varphi_0 \sim  {m_0^2/g_5 f_0}
%$.
For a scalar hadron $H$ created by an operator
of dimension $\Delta\geq1$, the integrand of \Eref{eq:THO}, 
$\mu_H \varphi_0 \sigma_H\sim z^{2\Delta-1}$, is small 
at small $z$.  Thus the integral is dominated by 
large $z\sim z_{max}$,
where the slowly-varying function $\varphi_0$ is 
of order $\hat\varphi_0$.
Therefore, using Eqs.~\eref{Hnorm}, \eref{eq:THO},
\eref{rhonorm} and \eref{fm2},
\bel{hereweare}
{f_0g_{0HH}\over m_0^2}\sim {f_0g_5 \hat\varphi_0\over m_0^2}  \int\ dz\ \mu_H(z) \sigma_H(z) \sim 1 \ .
\ee

 This approximate but generic relation applies to all $H$; 
it can fail for individual
hadrons or in extreme models, but will generally hold.
This explains the observations
in \cite{haduniv}.  
%Note that two extreme limits --- (1) when $H$ is
%created by a high-dimension operator (so that $\phi_H\approx
%\delta(z-\tilde z)$, for some $H$-independent $\tilde z\sim z_{max}$)
%and (2) when $H$ has a large radial excitation (so that $\phi_H$
%oscillates, with $H$-independent amplitude, down to small $z$) --- can
%easily be seen, from the above estimates, to be of the same order.
%This was observed but not explained in \cite{haduniv}.
In no limit, according to this argument, is \rcu\ exact across
the entire theory.
Instead, our estimates show that approximate \rcu\ is a general
property to be expected of all $\rho$ mesons in all AdS/QCD models.
Strictly we have only shown this for scalar hadrons, but the argument
is easily extended to higher spin hadrons.

These estimates are invalid when the $\rho$ meson mode is replaced
with the mode for any other hadron, unless that hadron is also the
lowest mode (structureless and positive-definite) created by a
conserved current (with model-independent normalization and boundary
condition at $z=z_{max}$.)  The only other hadrons 
of this type are the lowest spin-two glueball created by the
energy-momentum tensor and (if present) the lightest spin-3/2 hadron
created by the supersymmetry current. 
\begin{figure}[htbp]
  \begin{center}
    \leavevmode
     \epsfxsize=2.5in
     \hskip -1in \epsfbox{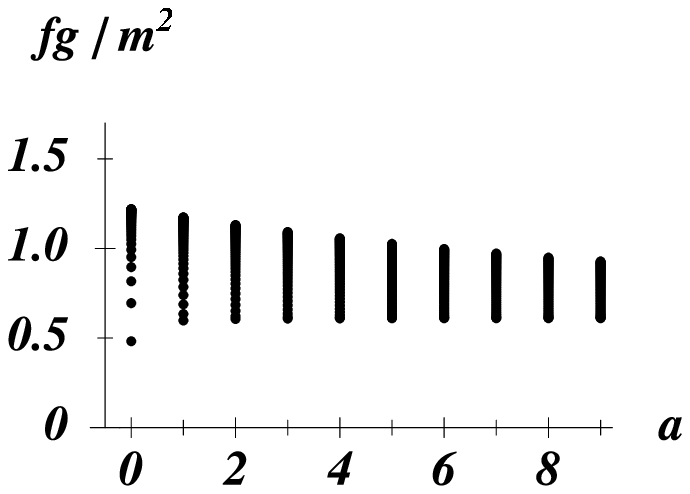}
  \end{center}
  \vskip -0.25in  \caption{The combination
 $f_0g_{0aa}^\Delta/m_0^2$ in the hardwall model,
plotted as a function of $a$ for all $\Delta<40$.}
  \label{fig:hardwall}
\end{figure}
\begin{figure}[htbp]
  \begin{center}
    \leavevmode
     \epsfxsize=2.5in
    \hskip -1in  \epsfbox{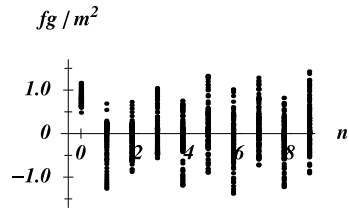}
  \end{center}
  \vskip -0.25in  \caption{The combination
 $f_n g_{naa}^\Delta/m_n^2$ in the
hardwall model, plotted as a function of $n$
for all $a$ and $\Delta<40$.}
  \label{fig:hardwallb}
\end{figure}

Now let us see that this argument applies in the hard-wall model and
D3/D7 model \cite{Myers,SonSteph}, reviewed in the appendices of
\cite{haduniv}.  (These should be viewed as 
toy models, as neither is precisely dual to a
confining gauge theory.
Calculations in specific gauge theories, such as the
duality cascade \cite{ikms}, are often straightforward but cannot
be done analytically.)
%The $\rho$ meson for the $SO(6)$ R-current in the
%hard-wall model is
%
%\be\label{Ahadwave} \varphi_0(z) = \frac{\sqrt{2}
%  z/z_{max} J_1 ( \zeta_{0;0} z/z_{max})}{R^3 J_1 ( \zeta_{0;0} )} 
% \ \ (J_0(\zeta_{0;0})=0 ) \ , \ee
%
%for which $\zeta_{0,0}$ is the first zero of $J_0(x)$.  Here
%$\hat z=z_{max}$ and $\varphi_0 = 1/R^3$, $\hat \mu_0 = R/z_{max}$, 
%$m0^2/f_0=0.624 (2\pi)^5/2N$.
%The $\rho$ meson for the flavor current in the D3/D7 model is 
%\begin{eqnarray*}
%\varphi_0= \phi^{II}_{0 ,0} & = & (\sqrt{12}/R^2) 
% (z/z_{max})^2 P_0^{(1,1)} ( 2 (z^2/z_{max}^2) - 1 )\ ,
%\end{eqnarray*}
%for which $\zeta_{0,0}$ is the first zero of $J_0(x)$.  Here
%$\hat z=$ and $\varphi_0 = $???, $\hat \mu_0 = ???$, $f_0/m_0^2=$.
Our statements about the $\rho$ meson modes in these models can be
checked using \cite{haduniv}, while those regarding the couplings
$g_{0 HH}$ are illustrated in the figures below.  Figure
\ref{fig:hardwall} shows the $\rho$'s couplings $g_{naa}^\Delta$ to
the scalar states $|\Delta,a\rangle$ of the hard-wall model, as a
function of the excitation level $a$, for all $\Delta<40$. As
expected, the couplings lie in a narrow range near 1.  That this is
true only of the $\rho$ is indicated in Fig.~\ref{fig:hardwallb},
where the couplings of all hadrons to the $n^{th}$ vector meson are
shown.  Only the $\rho$'s couplings lie in a narrow range near 1 and are
always positive.  These properties are also true in the D3/D7 model,
shown in the next two figures (with the additional feature \cite{haduniv}
 that the large-$\Delta$ limit and large-$a$ limit
of $g_{0aa}^\Delta$ are equal.)
\begin{figure}[htbp]
  \begin{center}
    \leavevmode
     \epsfxsize=2.5in
     \hskip -1in \epsfbox{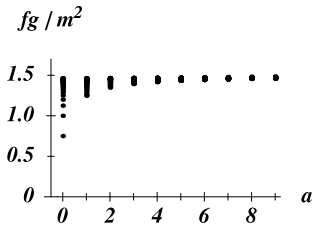}
  \end{center}
  \vskip -0.25in  \caption{As in Fig.~\ref{fig:hardwall}, for the D3/D7 model.}
%{The combination
% $f_0 g_{0aa}^\Delta/m_0^2$ in the D3/D7 model, plotted as a function of $a$
%for all $\Delta$.}
  \label{fig:D3D7}
\end{figure}
\begin{figure}[htbp]
  \begin{center}
    \leavevmode
     \epsfxsize=2.5in
     \hskip -1in \epsfbox{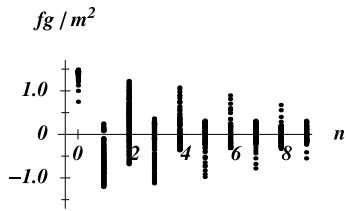}
  \end{center}
 \vskip -0.25in  \caption{As in Fig.~\ref{fig:hardwallb}, for the D3/D7 model.}
%  \caption{The combination
% $f_n g_{naa}^\Delta/m_n^2$ in the
% D3/D7 model, plotted as a function of $n$
%for all $a$ and $\Delta$.}
  \label{fig:D3D7b}
\end{figure}

\begin{figure}[htbp]
  \begin{center}
    \leavevmode
     \epsfxsize=2.5in
    \hskip -1in \epsfbox{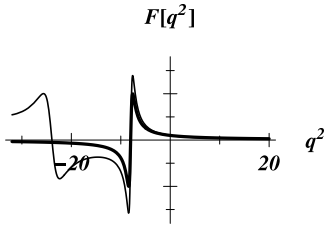}
  \end{center}
  \vskip -0.25in  \caption{$F(q^2)$ for the $|2,9\rangle$
state of the D3/D7 model; here $q^2$ has a small constant imaginary
part.  A fit (thick line) 
with a single pole misleadingly supports
\rd\ at small $|q^2|$.}
  \label{fig:fit}
\end{figure}

As is clear from the figures,
both models include hadrons for which
\rcu\ is approximately true but \rd\ fails badly.  
For instance, for the $|2,9\rangle$ state in the 
D3/D7 model, $f_0g^{(2)}_{0,9,9}/m_0^2\sim 1$,
but the terms in \Eref{Fnormrd} fall off slowly:
for $n=0,1,2,3,4\dots$
\be
f_n g_{n,9,9}^{(2)}/ m_n^2= 1.49 ,\ -1.21 ,\ 1.23  ,\ -1.12 ,\ 1.08 , \dots 
\ee
Yet a naive test (Fig.~\ref{fig:fit}) 
of \Eref{rddefn} moderately supports \rd.
Thus \rcu\ can lead to {\it apparent} \rd\
even if \rd\ is false.  This might be relevant for the 
success
of the \rd\ conjecture in QCD, which cannot be directly checked without
measuring the higher $g_{nHH}$.

The 
exploration of a toy scenario provides some additional,
though limited, perspective; it emphasizes the role of the
mass spectrum of the vector mesons.  For a scalar hadron created by an
operator of dimension $\Delta$, the condition of ultraviolet conformal
invariance implies $F(q^2)\to q^{-2(\Delta-1)}$.  This requires that
{\it at least} $\Delta-1$ of the terms in the sum be
non-vanishing. Suppose the sum is ``maximally truncated'': {\it only}
the first $\Delta-1$ couplings $g_{nHH}$, $n=0,\dots,\Delta-2$, are
non-vanishing.  Then the condition that $F(0)=1$ uniquely fixes the
coefficients in the sum:
\bel{truncfgm}
{f_n g_{nHH}\over m_n^2} = \prod_{k\neq n}^{\Delta-2} {m_k^2 \over m_k^2-m_n^2
}
\ee
The pattern of couplings thus depends on the pattern of
masses; both \rd\ and \rcu\ can fail, and
\rcu\ can be true even when \rd\ is false.  If,
for large $k$, $m_k
\sim k^{p}$ with $p\leq 1/2$, then \rcu\ fails at large
$\Delta$.  A flat-space stringy spectrum
$m_k\sim\sqrt{k}$ is a borderline case, while the low supergravity modes
of AdS/QCD have $p=1$.  Meanwhile examples in which \rd\ breaks down
are easily found.  If $m_k\sim (k+1)m_0$, then $|f_ng_{nHH}/m_n^2|$ is
largest for $n=0$ but remains large for moderate $n$.  If $m_k\sim
\sqrt{(k+2)(k+1)/2}\ m_0$, as in the D3/D7 model, then,
for large $\Delta$,
$|f_ng_{nHH}/m_n^2|$ is largest for $n> 0$ and \rd\ is badly
violated.  Some spin-one modes in the D3/D7 model are maximally
truncated, including the $\rho$, and for them \rd\ indeed fails.

\rem{Could our argument for generic and approximate \rcu\
be generalized to QCD?
The AdS/QCD relation provides some hope.  There is considerable
evidence from Regge physics \cite{Regge} that a fairly simple-minded
translation can be made from $k_\perp^2$ in gauge theory to $1/z^2$ in
string theory.  With proper definition of wave-functions of
$|k_\perp|$ (using operator matrix elements) for the $\rho$ and other
hadrons, one could imagine carrying the above argument into QCD.
}

In sum, neither \rd, nor the hypothesis that the $\rho$ is somehow a
four-dimensional gauge boson, are logically necessary for \rcu,
exact or approximate.  An independent AdS/QCD argument, applicable only to the
lightest mesons created by conserved currents, ensures the $\rho$'s
couplings are quasi-universal at large $\lambda$.  Could this argument
be generalized to QCD?  With suitable definition of hadronic 
wave-functions, using operator matrix elements, it might be possible to
directly extend this line of thinking to theories at arbitrary $\lambda$,
including QCD.  Alternatively, the applicability of our arguments
at smaller $\lambda$ could be tested using numerical
simulations of large-$N$ (quenched) QCD-like theories.

\

We thank D.T.~Son and L.G.~Yaffe for
useful conversations.  This work was supported by U.S. Department of
Energy grants DE-FG02-96ER40956 and DOE-FG02-95ER40893, and by an
 Alfred P. Sloan Foundation award.


\begin{thebibliography}{}

\bibitem{oldwork}
F.~Klingl, N.~Kaiser and W.~Weise,
%``Effective Lagrangian approach to vector mesons, their structure and decays,''
Z.\ Phys.\ A {\bf 356}, 193 (1996)
[arXiv:hep-ph/9607431];
%%CITATION = HEP-PH 9607431;%%
J.~J.~Sakurai,
%``Eight Ways of Determining the {\it rho} -Meson Coupling Constant''
Phys.\ Rev.\ Lett.\ {\bf 17}, 1021 (1966);
%%CITATION = PRLTA,17,1021;%%
J.~J.~Sakurai,
%``Theory Of Strong Interactions,''
Annals Phys.\  {\bf 11}, 1 (1960);
%%CITATION = APNYA,11,1;%%
M.~Gell-Mann and F.~Zachariasen,
%``Form-Factors And Vector Mesons,''
Phys.\ Rev.\  {\bf 124}, 953 (1961);
%%CITATION = PHRVA,124,953;%%
J.~J.~Sakurai,
Currents and Mesons 
(University of Chicago Press, Chicago, 1969);
\rem{
\bibitem{williams}}
for a review, see H.~B.~O'Connell, B.~C.~Pearce, A.~W.~Thomas and A.~G.~Williams,
%``Rho - omega mixing, vector meson dominance and the pion form-factor,''
Prog.\ Part.\ Nucl.\ Phys.\  {\bf 39}, 201 (1997)
[arXiv:hep-ph/9501251];
%%CITATION = HEP-PH 9501251;%%
%\cite{Williams:1997mg}
%% \bibitem{williams2}
A.~G.~Williams,
%``New results in vector meson dominance and rho meson physics,''
arXiv:hep-ph/9712405.
%%CITATION = HEP-PH 9712405;%%


%\cite{Bando:1987br}
\bibitem{hls}
For a review see M.~Bando, T.~Kugo and K.~Yamawaki,
%``Nonlinear Realization And Hidden Local Symmetries,''
Phys.\ Rept.\  {\bf 164}, 217 (1988).
%%CITATION = PRPLC,164,217;%%

%\cite{Harada:2003jx}
\bibitem{vm}
For a review see
M.~Harada and K.~Yamawaki,
%``Hidden local symmetry at loop: A new perspective of composite gauge boson
%and chiral phase transition,''
Phys.\ Rept.\  {\bf 381}, 1 (2003)
[arXiv:hep-ph/0302103].
%%CITATION = HEP-PH 0302103;%%


\bibitem{SonSteph}
D.~T.~Son and M.~A.~Stephanov,
%``QCD and dimensional deconstruction,''
Phys.\ Rev.\ D {\bf 69}, 065020 (2004)
[arXiv:hep-ph/0304182].
%%CITATION = HEP-PH 0304182;%%

\bibitem{AdS/CFT}
J.~M.~Maldacena,
%``The large N limit of superconformal field theories and supergravity,''
Adv.\ Theor.\ Math.\ Phys.\  {\bf 2}, 231 (1998)
[Int.\ J.\ Theor.\ Phys.\  {\bf 38}, 1113 (1999)]
[arXiv:hep-th/9711200];
%%CITATION = HEP-TH 9711200;%%
for a review, see
O.~Aharony, S.~S.~Gubser, J.~M.~Maldacena, H.~Ooguri and Y.~Oz,
%``Large N field theories, string theory and gravity,''
Phys.\ Rept.\  {\bf 323}, 183 (2000)
[arXiv:hep-th/9905111].
%%CITATION = HEP-TH 9905111;%%


\bibitem{haduniv}
S.~Hong, S.~Yoon and M.~J.~Strassler,
%``Quarkonium from the fifth dimension,''
arXiv:hep-th/0312071;
%%CITATION = HEP-TH 0312071;%%
%%``On the couplings of vector mesons in AdS/QCD,''
arXiv:hep-th/0409118.
%%CITATION = HEP-TH 0409118;%%

\bibitem{witten}
E.~Witten,
%``Anti-de Sitter space, thermal phase transition, and confinement in  gauge
%theories,''
Adv.\ Theor.\ Math.\ Phys.\  {\bf 2}, 505 (1998)
[arXiv:hep-th/9803131];
%%CITATION = HEP-TH 9803131;%%
S.~J.~Rey, S.~Theisen and J.~T.~Yee,
%``Wilson-Polyakov loop at finite temperature in large N gauge theory and
%anti-de Sitter supergravity,''
Nucl.\ Phys.\ B {\bf 527}, 171 (1998)
[arXiv:hep-th/9803135];
%%CITATION = HEP-TH 9803135;%%
A.~Brandhuber, N.~Itzhaki, J.~Sonnenschein and S.~Yankielowicz,
%``Wilson loops, confinement, and phase transitions in large N gauge  theories
%from supergravity,''
JHEP {\bf 9806}, 001 (1998)
[arXiv:hep-th/9803263].
%%CITATION = HEP-TH 9803263;%%

\bibitem{ikms}
J.~Polchinski and M.~J.~Strassler,
%``The string dual of a confining four-dimensional gauge theory,''
arXiv:hep-th/0003136;
%%CITATION = HEP-TH 0003136;%%
I.~R.~Klebanov and M.~J.~Strassler,
%``Supergravity and a confining gauge theory: Duality cascades and $\chi$SB-resolution of naked singularities,''
JHEP {\bf 0008}, 052 (2000)
[arXiv:hep-th/0007191].
%%CITATION = HEP-TH 0007191;%%



\bibitem{DIS}J.~Polchinski and M.~J.~Strassler,
%``Deep inelastic scattering and gauge/string duality,''
JHEP {\bf 0305}, 012 (2003)
[arXiv:hep-th/0209211].
%%CITATION = HEP-TH 0209211;%%

\bibitem{Myers}
A.~Karch and E.~Katz,
%``Adding flavor to AdS/CFT,''
JHEP {\bf 0206}, 043 (2002)
[arXiv:hep-th/0205236];
%%CITATION = HEP-TH 0205236;%%
M.~Kruczenski, D.~Mateos, R.~C.~Myers and D.~J.~Winters,
%``Meson spectroscopy in AdS/CFT with flavour,''
JHEP {\bf 0307}, 049 (2003)
[arXiv:hep-th/0304032].
%%CITATION = HEP-TH 0304032;%%







\end{thebibliography}
\end{document}